\documentclass[11pt]{article}

\usepackage[margin=1in]{geometry}
\usepackage{graphicx}
\usepackage{authblk}
\usepackage{amsmath,amssymb}
\usepackage{array}
\usepackage{caption}
\usepackage{xcolor}
\usepackage[normalem]{ulem}
\usepackage[numbers,sort&compress]{natbib}
\usepackage{hyperref}
\hypersetup{colorlinks=true, linkcolor=blue, citecolor=blue, urlcolor=blue}

\newif\ifshowrevisions
\showrevisionsfalse          
\newcommand{\red}[1]{\ifshowrevisions\textcolor{red}{#1}\else#1\fi}

\begin{document}

\title{Metasurfaces for neutral-atom trapping}

\author[1]{Chengyu~Fang}
\author[1]{Minjeong~Kim}
\author[2,3]{Mark~Saffman}
\author[1,2]{Jennifer~T.~Choy}
\author[1,2,*]{Mikhail~A.~Kats}

\affil[1]{Department of Electrical and Computer Engineering, University of Wisconsin-Madison, 1415 Engineering Drive, Madison, WI 53706, USA}
\affil[2]{Department of Physics, University of Wisconsin-Madison, 1150 University Avenue, Madison, WI 53706, USA}
\affil[3]{Infleqtion, Madison, WI 53703, USA}
\affil[*]{Corresponding author: \texttt{mkats@wisc.edu}}

\date{}

\maketitle
\vspace{-3.5em}

\begin{abstract}
Trapped neutral atoms are one of the leading platforms for quantum information technologies, in particular for quantum computing, but scaling them to array sizes needed for utility-scale quantum computing is a major engineering challenge. Here we review optical metasurfaces as an enabling technology that provides fine control over the phase, amplitude, and polarization of light, with pixel counts far exceeding what is available with spatial light modulators (SLMs) and other active devices. The large pixel counts have recently led to demonstrations of static arrays of optical tweezers with hundreds of thousands of sites and arrays of optical \red{bottle beams} with complex three-dimensional trapping profiles. The flexibility and scalability of optical metasurfaces provides a route towards miniaturized, integrated, and highly scalable atomic experiments and instruments.
\end{abstract}

\section{Introduction}

Neutral-atom arrays are enabling components for a class of quantum computers, and for quantum networking and sensing \cite{Kaufman2021}. The primary existing approach to trap arrays of atoms from clouds of cold atomic vapor involves the use of active optical devices, such as spatial light modulators (SLMs) \cite{Nogrette2014, Kim2016, Barredo2016, Manetsch2025}, digital micromirror devices (DMDs) \cite{Stuart2018, Wang2020}, and acousto-optic deflectors (AODs) \cite{Endres2016, Bluvstein2022, Bluvstein2024} to form arrays of optical dipole traps. These active devices can generate dipole-trap arrays with thousands of sites \cite{Manetsch2025}, enable atom rearrangement to form defect-free arrays \cite{Kim2016}, and facilitate coherent transport between sites for non-local connectivity \cite{Chiu2025, Bluvstein2024}.

Despite the flexibility offered by active optical devices, further scaling and integration of neutral-atom arrays face several engineering challenges. For example, the pixel count of SLMs and DMDs limits the number of traps a single device can generate, and a similar limitation exists in AODs related to the radio-frequency bandwidth. It is possible to use multiple active devices simultaneously to get around, e.g., the pixel limitations of SLMs \cite{Singh2022, Manetsch2025}, but the integration is complex, and so far the number of traps generated via SLMs (up to 12,000 sites \cite{Manetsch2025}) falls short of the predicted number of qubits needed to achieve quantum advantage in factoring integers \cite{Gidney2025, Zhou2025} or quantum chemistry problems \cite{Lee2021, Alexeev2024}. We note that a recent work argues that the requirement could be brought down to as few as $\sim$10{,}000 reconfigurable atomic qubits for Shor's algorithm \cite{Cain2026}, albeit with runtimes ranging from months to over a century at that qubit count; practical runtimes of days to months still require roughly an order of magnitude more physical qubits.

The limitations of active optical devices for atom trapping have motivated the design of trap arrays using  passive optical elements, including micro-lens arrays (MLAs) \cite{Dumke2002, Ohl2019, Schlosser2023, Pause2023, Pause2024} and re-imaged optical intensity masks \cite{Huft2022, Fang2025}. These passive components can be manufactured to large scales, potentially enabling very large atom arrays. However, both of these technologies need extra optics to deliver and de-magnify the beams into the vacuum cell, hindering integration and miniaturization. These approaches also have limited functionality; for example, MLAs are not able to easily form complex beams such as optical bottles. We note that optical lattices can generate a large number of trapping sites \cite{Park2022,Tao2024,Gyger2024}, but their typical sub-micron site spacing makes addressing individual atoms difficult and incompatible with Rydberg gates.

Here we review an emerging approach to creating large arrays of optical traps for neutral-atom quantum computing, using optical metasurfaces manufactured using semiconductor fabrication techniques. Metasurfaces are flat optical components that provide subwavelength control of optical wavefronts, and can replace combinations of active and passive optical components \cite{Yu2011}. We discuss how metasurfaces may enable (i) the scaling of optical-tweezer arrays to millions of sites, (ii) the generation of complex trapping profiles like optical bottle beams, (iii) multiple functionalities like polarization multiplexing or trapping and imaging in a single optic. We also discuss practical aspects and challenges of metasurface design, including material choices, miniaturization, and integration into complex optical and vacuum systems.

\section{Array scalability}
\label{scalability}

\begin{figure}[!ht]
  \centering
  \includegraphics[width=\linewidth]{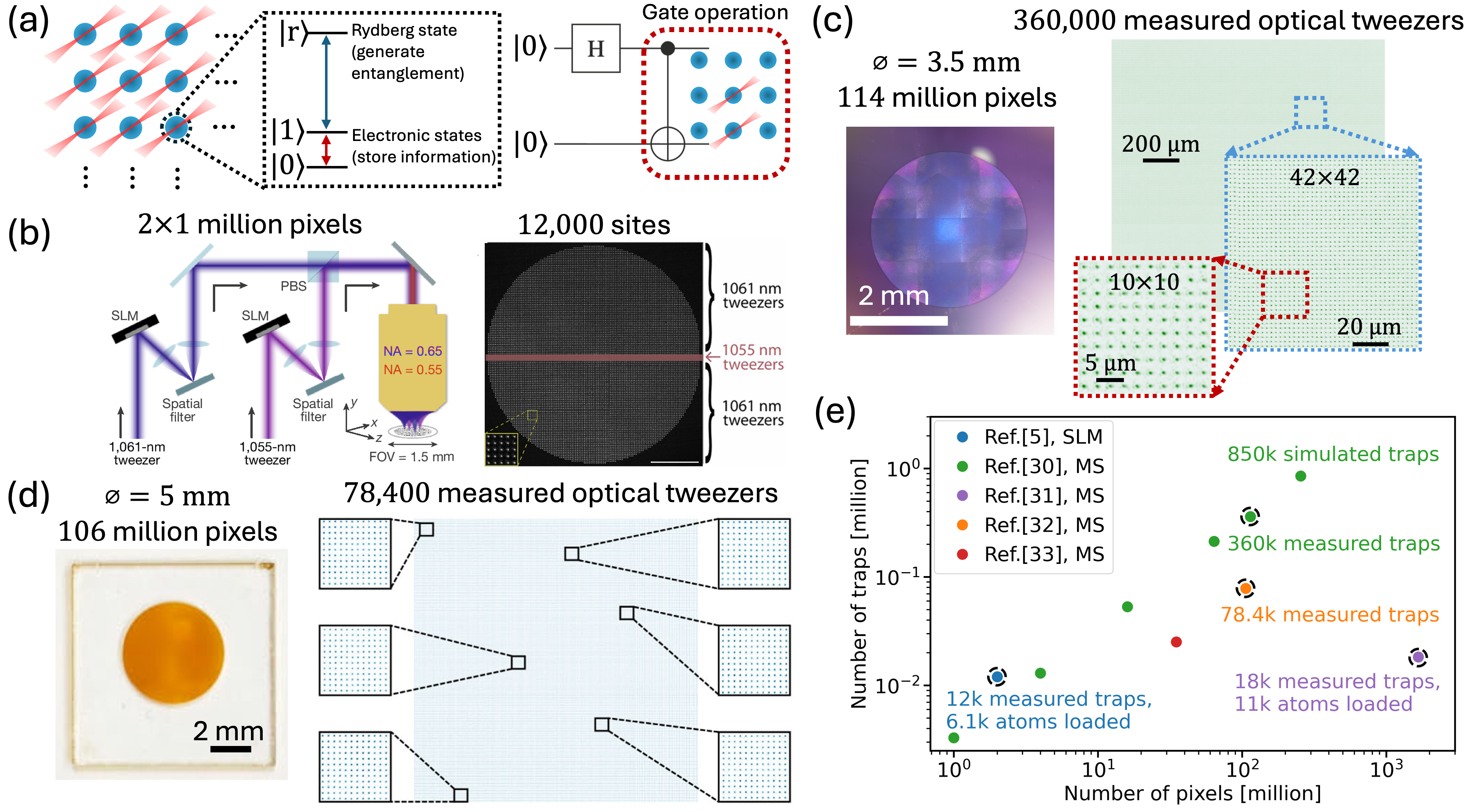}
  \caption{Approaches to scale neutral-atom tweezer arrays. (a) Schematic of a neutral-atom quantum computer, where atoms are trapped in an optical-tweezer array, with qubits encoded in two electronic states ($|0\rangle$ and $|1\rangle$). A Rydberg state ($|r\rangle$) enables entanglement between neighboring atoms, enabling multi-qubit gates such as the CNOT shown on the right. (b) An array of 12,000 tweezer sites generated by two $1024 \times 1024$ SLMs ($\sim$2 million pixels in total) using 1061~nm and 1055~nm \red{lasers}, with 6,100 atoms loaded; reproduced from \cite{Manetsch2025}. (c) An array of 360,000 tweezer sites generated by a 3.5-mm-diameter metasurface with 114 million pixels (no atoms loaded in the device shown, although single-atom trapping was demonstrated in a smaller array reported in \red{the same} work); adapted from \cite{Holman2026}. (d) An array of 78,400 tweezer sites generated by a single 5-mm-diameter metasurface with 106 million pixels (no atoms loaded); adapted from \cite{Wang2025}. (e) Summary of literature on metasurfaces for atom trapping \cite{Wang2025, Holman2026, Fang2026, Wang2026}, with the number of tweezers formed as a function of the number of metasurface pixels. For reference, the record result using two SLMs is also included \cite{Manetsch2025}. Dashed circles mark \red{experimental demonstrations}; the remaining data points correspond to simulated designs. ``Measured traps'' means the traps were imaged, but atoms may not have been loaded.}
  \label{fig:figure1}
\end{figure}

Scalability of trapped-atom arrays is essential for advancing quantum technologies, especially for quantum computing. In a Rydberg-atom-based quantum computer (Fig.~\ref{fig:figure1}(a)), each trapped atom stores quantum information in two electronic states, which are hyperfine ground states for alkali atoms such as Rb and Cs, or nuclear-spin states for alkaline-earth(-like) atoms such as Sr and Yb \cite{Kaufman2021}. Neighboring atoms can be entangled after being excited to Rydberg states \cite{Saffman2010}. Quantum gate operations are then implemented with sequences of laser or microwave pulses applied either globally to the array or to individual atoms. Solving complex quantum chemistry problems, such as ground-state energy estimation of iron-molybdenum cofactor (FeMoco), requires millions of physical qubits \cite{Alexeev2024}, because thousands of logical qubits are required and each logical qubit may require up to thousands of physical qubits \cite{Zhou2025}.

\begin{table}[!t]
    \small
    \centering
    \caption{Comparison of optical devices for generating tweezer arrays}
    \newcolumntype{C}[1]{>{\centering\arraybackslash}m{#1}}
    \newcolumntype{T}[1]{>{\centering\arraybackslash}p{#1}}
    \renewcommand{\arraystretch}{1.9}
    \begin{tabular}{C{2.2cm}C{1.8cm}C{3.2cm}C{3.2cm}}
        \hline
        \multicolumn{1}{C{2.2cm}}{Device} & \multicolumn{1}{C{1.8cm}}{Refresh rate} & \multicolumn{1}{C{3.2cm}}{Largest number of optical traps demonstrated} & \multicolumn{1}{C{3.2cm}}{Largest number of single atoms loaded} \\
        \hline
        SLM            & sub kHz      & $\sim$12,000 sites \cite{Manetsch2025}  & $\sim$6,100 atoms \cite{Manetsch2025}  \\
        DMD            & $\sim$ kHz   & 20 sites \cite{Stuart2018}        & $\sim$2 atoms \cite{Stuart2018}        \\
        AOD            & $\sim$ MHz   & \red{1,440 sites \cite{Chiu2025}}        & \red{$\sim$720 atoms \cite{Chiu2025}}       \\
        Amplitude mask & static       & 1,225 sites \cite{Huft2022}          & $\sim$370 atoms \cite{Huft2022}        \\
        MLA            & static       & $\sim$3,000 sites \cite{Pause2024}         & $\sim$1167 atoms  \cite{Pause2024}     \\
        Metasurface    & static       & 360,000 sites \cite{Holman2026}      & $\sim$11,000 atoms \cite{Wang2026}      \\
        \hline
    \end{tabular}
    \label{tab:table1}
\end{table}

Here, we compare the experimentally demonstrated active and passive optical devices to generate tweezer arrays (Table~\ref{tab:table1}). We focus on how each approach scales with the number of trap sites, and what (if anything) will limit further scalability. SLMs used in atom-trapping experiments are typically liquid-crystal phase-only panels that apply a pixel-by-pixel $0$--$2\pi$ phase profile to the incident laser beam; then a high-NA lens focuses the resulting wavefront into the vacuum cell. The phase profile at the SLM is typically calculated using the Gerchberg-Saxton (G\mbox{-}S) algorithm \cite{Gerchberg1972} or its variants \cite{Tao2015,Fang2026}, to achieve some target trapping profile. This phase-only (holographic) approach redistributes the energy of light in a reasonably power-efficient manner, especially for sparse arrays, with typical first-order diffraction efficiency of $40-60\%$ \cite{Kim2016,Nishimura2026} (meaning about half of the light is lost). Due to the relatively slow response of liquid crystals (typically tens to hundreds of Hz), SLMs are primarily used for generating static traps, although  rearrangement to achieve defect-free arrays using SLMs has also been demonstrated \cite{Kim2016, Lin2025}.

Until very recently, the largest experimentally demonstrated atom array \cite{Manetsch2025} used two $1024 \times 1024$ SLMs in parallel to achieve one complete array with $\sim$12,000 trapping sites, stochastically loaded with $\sim$6,100 atoms (Fig.~\ref{fig:figure1}(b)). In such experiments, each site is loaded with a single atom with near 50\% probability, although enhanced-loading techniques can raise the loading rate to near 90\% \cite{Kaufman2021}. To further scale the number of traps toward the million-site level, the SLM approach faces a bottleneck in the pixel budget. With a typical ratio of $\sim$300 pixels per trap \cite{Holman2026}, a million-site array would require hundreds of millions of pixels, which are orders of magnitude beyond the available pixel counts in commercial SLMs. The gap is even larger for the high-power variants of SLMs, which offer fewer pixels due to tighter thermal constraints, but are often required because trapping a single atom typically requires mW-level laser power, which scales to hundreds of watts for a large array. As an example, Ref.~\cite{Manetsch2025} uses total laser power exceeding 100 watts. SLMs also generally have lower diffraction efficiency when illuminated by high-power lasers due to the degradation of performance in the liquid crystals \cite{Wang2023}, which further hinders the scalability.

The DMD, another pixelated optical device, imprints a binary amplitude mask on the beam, with grayscale amplitude produced by kilohertz (kHz) pulse-width modulation (PWM) on each micro-mirror. The beam profile can then be re-imaged into the vacuum cell to trap atoms. Because DMDs act as amplitude masks, they are less power-efficient than phase SLMs for sparse tweezer arrays \cite{Zupancic2016}, and their PWM modulation can drive parametric heating when it overlaps with the atom trapping frequencies \cite{Stuart2018}. Due to these limitations, DMDs are not commonly used for single-atom array generation.

AODs deflect the incident beam through Bragg diffraction from an acoustic wave in a crystal driven by radio-frequency (RF) tones, where applying multiple tones simultaneously can generate several beams at different angles, followed by focusing to generate a one-dimensional array of traps. A two-dimensional array can be generated by two orthogonally positioned AODs. AODs operate at megahertz (MHz) rates which enables fast movement of single atoms \cite{Endres2016, Bluvstein2022}. However, the number of independently addressable beams is set by the RF bandwidth divided by the minimum RF tone spacing, typically limiting a single AOD to tens of beams. So, AODs are best suited for fast reconfiguration instead of generating large static arrays. These complementary trade-offs to SLMs motivate the hybrid architecture of using SLMs for generating large static arrays and AODs for fast rearrangement of atoms between sites \cite{Bluvstein2024, Manetsch2025}.

Active devices provide the capabilities to rearrange atoms \cite{Endres2016, Barredo2016, Lin2025} and improve the uniformity by using real-time feedback \cite{Chew2024}. However, they face significant engineering challenges to scale to much larger atom numbers. SLMs are the leading platform to generate large-scale trap arrays, but they still do not provide a clear path to scale to millions of trap sites. A solution is passive optical components that are simpler to design and fabricate than a complete SLM, and can thus be easily scaled. For example, a laser beam incident on a passive optical intensity mask, e.g., an array of apertures, can be turned into an array of optical traps via re-imaging of the mask surface into a vacuum cell. Binary \cite{Huft2022} and ternary \cite{Fang2025} intensity masks are straightforward to make up to arbitrary sizes using conventional fabrication processes, and spatial filtering in the Fourier plane can yield high-quality optical traps. Amplitude masks have been used to demonstrate arrays of tweezer beams, bottle beams, and interleaved arrays that can trap individual atoms of two species using a single laser \cite{Fang2025}. Because the re-imaging optics can be designed to have various demagnifications, the feature size on the mask can be relatively large and easily achievable by even low-resolution lithography, which makes fabrication easy and readily scalable. However, similarly to DMDs, the trapping approach using amplitude optical masks has limited power efficiency, especially for sparse tweezer arrays, where a large fraction of light ends up reflected or absorbed at the mask.

A micro-lens array (MLA) is another type of passive optic that can generate traps when illuminated by a trapping laser \cite{Dumke2002, Ohl2019, Schlosser2023, Pause2023, Pause2024}, with pitch sizes ranging from tens to hundreds of microns. Each micro-lens focuses the incident light into a spot, and this spot can be de-magnified and re-imaged into the vacuum cell for atom trapping. MLA-based tweezer arrays have been scaled to $\sim$3,000 sites \cite{Pause2024}. However, further scaling is limited by several factors: the size of each micro-lens limits the number of lenses that fit within a reasonable optical aperture of the imaging system, and the dead area between neighboring micro-lenses limits the power efficiency.

An approach that can generate very large optical-trap arrays is optical lattices, in which traps are formed by standing waves. Optical lattices can generate thousands to as many as $\sim$$10^6$ trapping sites \cite{Park2022,Tao2024,Gyger2024}, but the site spacing is typically sub-micron, which makes it difficult to address individual atoms and perform Rydberg gates. Several-micron spacings can be achieved in optical lattices using small-angle crossed beams, but the resulting arrays have been limited to a few hundred atoms \cite{Nelson2007, Wang2015}. In addition, the arrangement of the resulting traps is limited due to the nature of standing waves.

Perhaps the most flexible passive optical components are optical metasurfaces, which are planar optical elements that provide subwavelength control over the phase, amplitude, and/or polarization of light passing through them. Metasurfaces can serve a similar role to SLMs, generating arbitrary field profiles holographically, but they have much smaller pixels (i.e., metasurface unit cells), suppressing higher-order diffraction that limits efficiency in SLMs, and can be scaled to nearly arbitrary pixel counts, which are only limited by the writing time of high-resolution lithography. Two recent studies demonstrated the scaling opportunities offered by metasurfaces, with Holman et al. generating 360,000 tweezer sites using a 3.5-mm-diameter metasurface containing 114 million titanium-dioxide meta-atoms (Fig.~\ref{fig:figure1}(c)) \cite{Holman2026}, and Wang et al. demonstrating tweezer arrays with 78,400 sites using a 5-mm-diameter metasurface containing 106 million silicon-nitride meta-atoms (Fig.~\ref{fig:figure1}(d)) \cite{Wang2025}. Very recently, a single 19.8-mm-diameter metasurface with about 1.7 billion silicon-nitride meta-atoms trapped about 11,000 atoms in an 18,225-site array \cite{Wang2026}, exceeding the largest atom number reported with SLMs \cite{Manetsch2025}. Figure~\ref{fig:figure1}(e) shows that the number of traps can scale roughly linearly with pixel count, with the slope set by factors such as the target array uniformity \cite{Holman2026}. We note that the metasurface in Ref.~\cite{Wang2026} has 1.7 billion pixels which far exceeds what is needed for the demonstrated 18,225 traps; in that work, the larger pixel number was used to enlarge the aperture to create a longer working distance, enabling the metasurface positioned outside of the vacuum cell to generate traps inside the cell. The first-order diffraction efficiency of these metasurfaces was $60\%$ in Ref.~\cite{Holman2026} and $67.5\%$ in Ref.~\cite{Wang2026}, comparable to or exceeding those typical of SLMs \cite{Kim2016,Nishimura2026}.

Beyond the current demonstrations, the pixel budget of a single metasurface has a path to scale to much larger sizes. In most laboratory demonstrations, metasurfaces are patterned using electron-beam lithography \cite{Hsu2022,Huang2023,Huang2024,Wang2025,Holman2026,Fang2026}, which is a serial technique and thus limits the achievable pixel counts. However, other lithography methods, e.g., deep-ultraviolet \cite{Park2024} and nanoimprint \cite{Oh2021} lithography, can be used to dramatically improve the scalability and throughput. As an example, an all-glass 100-mm-diameter visible metalens with 18.7 billion pixels has been demonstrated \cite{Park2024}, which is one order of magnitude larger than the largest metasurface fabricated for atom trapping \cite{Wang2026}. 

We note that today's metasurfaces do not address all of the challenges in building very large uniform arrays. Beyond the need for sufficiently high laser power and large field of view of the optics, it is also not possible to achieve uniform, defect-free atomic arrays using purely static optical traps. Unlike SLMs, which can use closed-loop feedback to correct the trap non-uniformity \cite{Manetsch2025}, the phase profiles of metasurfaces are fixed at fabrication. Likely future approaches to improving uniformity of large arrays include fabrication of multiple slightly varying designs and selecting the most-uniform one, given a stable laser system, or introduction of some active fine-tuning mechanism.

\section{Generation of complex beam profiles}

The large pixel count of metasurfaces enables more sophisticated amplitude and phase control of a beam, enabling complex three-dimensional traps. In this section, we focus on using metasurfaces to generate complex trapping profiles, such as one-dimensional and two-dimensional dark-trap arrays and interleaved bright- and dark-trap arrays.

A key building block of these profiles is the optical bottle beam, which has a dark ``focal spot'' surrounded by a bright shell. Optical bottle beams can trap atoms at the intensity-null center when the laser is blue-detuned from the atomic transition. Bottle beams are more difficult to form than tweezers because they have a broader Fourier spectrum \cite{Pasienski2008}, and significantly better phase control in the plane of traps is required to form a complete three-dimensional bottle, requiring more metasurface pixels.

Despite the added complexity, optical bottles offer several advantages for neutral-atom applications, especially computing \cite{Saffman2010}. First, trapping atoms at an intensity null rather than at \red{an intensity} maximum reduces photon scattering from the trap light, supporting longer trap lifetimes and coherence times \cite{Li2012, Ozeri1999, Kuhr2005, Grimm2000}. Also, atoms in bottle beams are less sensitive to trapping-laser intensity fluctuations because the light intensity at the atom position is small, so the same fractional fluctuation produces a smaller perturbation compared to optical tweezers. Finally, bottle beams avoid workarounds needed because the AC polarizability of alkali atoms flips sign between ground and Rydberg states at typical tweezer wavelengths \cite{Saffman2005}, which results in anti-trapping of atoms in a tweezer during Rydberg excitation. Existing workarounds include using near-resonance trapping wavelengths that avoid anti-trapping but increase photon scattering \cite{Ahlheit2025, Saffman2005}, or switching off the trap during Rydberg excitation and recapturing the atom afterward \cite{Bernien2017}, which can cause atom loss and motional heating \cite{Wilk2010, Lorenz2021, De2023}. In a blue-detuned bottle beam, atoms remain confined during Rydberg excitation \cite{Barredo2020}, and a ``magic'' wavelength can further be chosen at which the polarizability is the same for ground and Rydberg states, eliminating differential light shifts \cite{Zhang2011}. 

For alkaline-earth(-like) species with two valence electrons, such as Sr and Yb, this anti-trapping problem can be circumvented in a different way \cite{Kaufman2021}. Rydberg states of these atoms have one excited electron, while the second valence electron remains bound to the ion core and provides strong optical transitions. At trap wavelengths red-detuned from these transitions, the bound valence electron contributes positive polarizability that can compensate the negative ponderomotive polarizability of the excited electron, so the ground and Rydberg states of these divalent atoms can be trapped in the same bright tweezers. That being said, dark traps are still attractive for these species, since atoms in dark traps have less photon scattering and reduced sensitivity to laser fluctuations \cite{Grimm2000,Li2012,Ozeri1999,Kuhr2005}.

\begin{figure}[!ht]
  \centering
  \includegraphics[width=\linewidth]{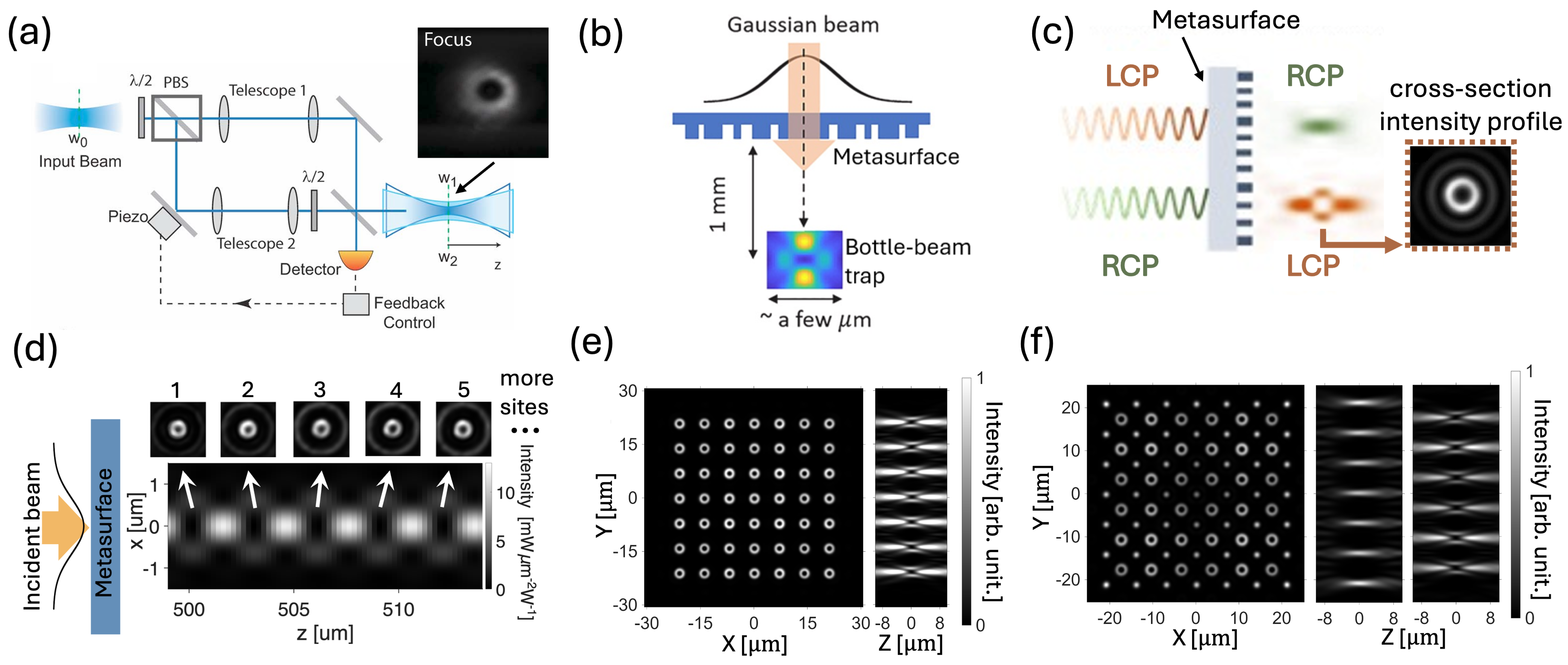}
  \caption{Generation of optical bottle beams. (a) Bottle-beam trap generated by destructive interference of two co-propagating Gaussian beams with different waists, stabilized by an active phase lock; adapted from \cite{Isenhower2009}. (b) Single dark trap generated from an incident Gaussian beam by a $\sim$1-mm metasurface, optimized for high efficiency and trap quality; adapted from \cite{Xiao2021}. (c) Polarization-multiplexed metasurface that generates a bright trap for one circular polarization and a dark trap for the opposite circular polarization; adapted from \cite{Cheng2025}. (d) One-dimensional array of ten dark traps generated by a metasurface; five of the sites are shown in X-Z and \red{X-Y}, for the optic axis along Z; adapted from \cite{Lim2023}. (e) Two-dimensional $7\times7$ dark-trap array generated by a single metasurface, shown in X-Y (in the focal plane; left) and in X-Z (right) along the trap centers; adapted from \cite{Fang2026}. (f) Two-dimensional array of interleaved $7\times7$ bright traps and $6\times6$ dark-trap arrays generated by a single metasurface, shown X-Y (left) and X-Z along the bright array (middle), and along the dark array (right); adapted from \cite{Fang2026}.}
  \label{fig:figure2}
\end{figure}

Several approaches have been demonstrated to form a single bottle beam. A bottle beam can be created via destructive interference between two Gaussian beams with slightly different waists (Fig.~\ref{fig:figure2}(a) \cite{Isenhower2009}), or using a phase-only SLM \cite{Barredo2020}. Single bottle beams have also been demonstrated using metasurfaces \cite{Xiao2021,Cheng2025}. For example, as shown in Fig.~\ref{fig:figure2}(b), a single passive metasurface illuminated by a Gaussian beam can generate a bottle-beam trap at the focal plane with higher efficiency compared to a two-beam interference approach. A polarization-multiplexed metasurface has also been used to generate a bottle beam (Fig.~\ref{fig:figure2}(c) \cite{Cheng2025}), with the polarization multiplexing functionality discussed in the following section.

Bottle beams are more difficult to form than optical tweezers because they require simultaneous control over both amplitude and phase in the focal plane to form the traps in three dimensions \cite{Fang2026}. This added degree of freedom compared to optical tweezers demands more pixels when the wavefront is generated holographically using SLMs or metasurfaces. For this reason, to our knowledge, there have not been demonstrations of large arrays of bottle beams using an SLM. Metasurfaces, which can provide orders-of-magnitude more pixels than SLMs (Section~\ref{scalability}), are therefore a natural route to large, high-quality bottle-beam arrays.

A one-dimensional array of ten optical bottles was recently implemented using a metasurface with TiO$_2$ meta-atoms, where each optical bottle is formed along the optic axis at locations of point phase singularities (Fig.~\ref{fig:figure2}(d)) \cite{Lim2023}. A two-dimensional array of bottle beams was demonstrated using a silicon-on-sapphire metasurface, where the array was in a single focal plane (Fig.~\ref{fig:figure2}(e) \cite{Fang2026}), which is the atom geometry most commonly used in neutral-atom quantum computing using SLMs. In Ref. \cite{Fang2026}, the conventional G-S algorithm for designing the phase profile on the metasurface was insufficient because it does not enforce a particular phase profile in the focal plane, and a modified G-S algorithm that enforces both amplitude and phase near the trapping sites (but leaves the field unconstrained elsewhere) was developed instead.

The modified G-S approach was also used to design metasurfaces that generate arrays in which bright and dark traps are spatially interleaved (Fig.~\ref{fig:figure2}(f) \cite{Fang2026}). By selecting a trapping wavelength at which two atomic species have polarizabilities of opposite signs, a single laser can be used to trap one atomic species in the bright trap, and the other in the dark trap, which can be advantageous for quantum error correction \cite{Singh2023}. More complicated two-dimensional geometries, such as interleaved dual-species pentagonal lattices \cite{Petrosyan2024}, can also be readily generated using metasurfaces.

\section{Multi-functionality, miniaturization, and integration}

Beyond the optics used to generate trap arrays, the complexity of neutral-atom hardware also includes vacuum components, infrastructure to cool atomic ensembles, and optics for fluorescence readout. Briefly, the process flow of a typical neutral-atom-array experiment has the following stages. First, a magneto-optical trap cools and collects an ensemble of atoms using multiple pairs of counter-propagating, circularly polarized beams that are aligned and balanced over a millimeter- to centimeter-scale volume \cite{Raab1987}. Individual atoms are then loaded from this cloud of atoms into an array of tightly focused dipole traps \cite{Grimm2000}, which usually requires high-numerical-aperture (high-NA) optics that form many diffraction-limited spots with uniform trap depth across a large field of view. Once loaded, the atoms are manipulated with additional laser and microwave fields for operations such as state preparation and Rydberg-mediated gates \cite{Saffman2010}, often addressing individual sites and therefore demanding fast, high-resolution beam control. Finally, the state of each atom is read out by collecting its fluorescence \cite{Nelson2007} through a high-NA lens onto a low-noise, high-sensitivity camera, where the requirements shift to high collection efficiency.

These various steps require many optical components, including non-standard ones modified for working with vacuum cells, such as specialized objectives for imaging through thick glass walls \cite{Li2020}, and anti-reflection coatings for high-NA optics. Each additional optical component introduces alignment degrees of freedom and can contribute to wavefront distortion and power loss. This becomes increasingly more difficult to manage for systems that need to either scale or to be compact and portable.

The metasurface approach can help reduce the number of distinct optical components. Because each meta-atom can be designed to have different responses as a function of wavelength and polarization, a single metasurface can perform multiple optical functions. In this section, we show how multi-functional metasurfaces are becoming incorporated in atom-trapping experiments, and how they can enable more-integrated atomic instruments.

\begin{figure}[!ht]
  \centering
  \includegraphics[width=\linewidth]{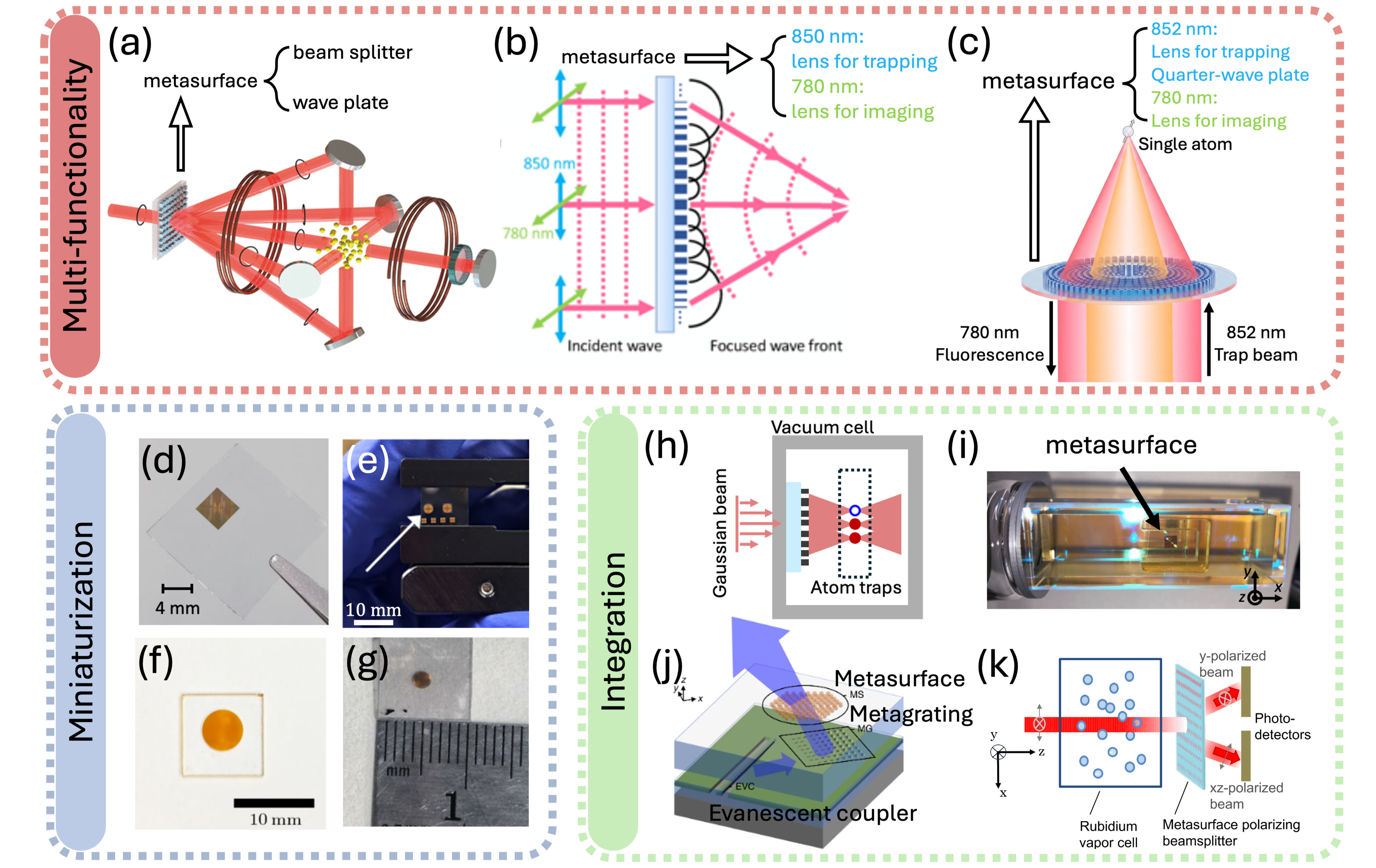}
  \caption{Multi-functionality, miniaturization, and integration enabled by metasurfaces in atom-trapping optical systems. (a) Multi-functional metasurface for generating a MOT, combining beam splitting and polarization control to generate five cooling beams from one incident beam; adapted from \cite{Jin2023}. (b) Polarization-multiplexed metalens, operating for two wavelengths (780 nm and 850 nm) for two orthogonal linear polarizations; adapted from \cite{Hsu2022}. (c) Wavelength-multiplexed metalens at 780 nm and 852 nm, which adds quarter-wave-plate functionality only at 852 nm; adapted from \cite{Chen2024}. (d--g) Representative compact metasurfaces that range from sub-millimeter to a few millimeters; adapted from \cite{Hsu2022,Fang2026,Wang2025,Chen2024}. (h) Schematic of bright- and dark-trap generation using a metasurface integrated inside a vacuum cell; adapted from \cite{Fang2026}. (i) An amorphous silicon on fused silica metasurface integrated inside a vacuum glass cell to replace \red{the high-NA objective}; adapted from \cite{Hsu2022}. (j) Integrated metasurface platform to launch a centimeter-scale circularly polarized beam for generating MOTs; adapted from \cite{Ropp2023}. (k) Metasurface integrated to a rubidium-vapor atomic magnetometer for balanced polarimetry; adapted from \cite{Yang2024}.}
  \label{fig:figure3}
\end{figure}

Before single atoms can be trapped in tweezer or bottle-beam traps, a cloud of atoms must be localized and cooled using a magneto-optical trap (MOT) \cite{Raab1987}, which conventionally uses three pairs of counter-propagating circularly polarized beams with controlled handedness. This typically requires multiple waveplates, polarizing beam splitters, and mirrors (often retroreflectors), with demanding alignments. Multi-functional metasurfaces provide a path to reduce the number of optics in a MOT. For example, Fig.~\ref{fig:figure3}(a) shows a metasurface that splits one circular-polarized beam into five output beams with uniform power ratios and circular polarizations, enabling a compact MOT \cite{Jin2023}. The beam-splitting function was achieved by using a gradient phase profile, while the wave-plate function was enabled by birefringent meta-atoms at each pixel. Building on this platform, Ref.~\cite{Tian2025} used a second metasurface to replace the wave plate and the mirror that remained in Fig.~\ref{fig:figure3}(a), further reducing the number of optics.

In addition to magneto-optical trapping and then dipole trapping of single atoms, atomic experiments need readout optics, achieved using fluorescence imaging \cite{Kaufman2021}. The imaging and trapping optical paths usually share one high-NA objective, but are then split using a dichroic into the trapping path (into the cell) and a path for fluorescence imaging (out of the cell, at the resonance wavelength of the atoms) leading to a single-photon camera. Metasurfaces can in principle streamline these optical systems, with initial demonstrations including a polarization-multiplexed metalens, with one polarization for the trapping wavelength and another for the imaging wavelength (Fig.~\ref{fig:figure3}(b)) \cite{Hsu2022}, and a wavelength-multiplexed metalens which builds in a wave-plate functionality at only the trapping wavelength (Fig.~\ref{fig:figure3}(c)) \cite{Chen2024}. In both of these demonstrations, efficiency remains a challenge because the multi-functionality was enabled by polarization multiplexing \cite{Hsu2022} or spatial interleaving \cite{Chen2024}, although potentially higher-efficiency achromatic designs can be achieved by engineering the dispersion of each meta-atom \cite{Chen2018}.

While providing multi-functionality, metasurfaces remain compact for two reasons. First, each meta-atom has subwavelength dimensions (hundreds of nanometers) ~\cite{Yu2011} instead of several to tens of micrometers per pixel in phase SLMs. Second, because they are passive, metasurfaces do not need the electronics and cooling hardware that surrounds SLM panels. This size difference allows a millimeter-scale metasurface to contain $10^7$--$10^8$ independently patterned pixels ~\cite{Hsu2022, Fang2026, Wang2025, Chen2024}, which is one to two orders of magnitude more than that of a bulky SLM.

Their compact size makes metasurfaces easier to integrate with other atom-trapping hardware, like vacuum cells. As proposed in Ref.~\cite{Fang2026}, a metasurface can be put directly into the vacuum cell (Fig.~\ref{fig:figure3}(h)), which can reduce the wavefront distortion introduced by the cell wall \cite{Pritchard2016}, and remove the need for sophisticated objectives that compensate for the wall thickness. The integration is possible because most metasurfaces comprise dielectrics and semiconductors which can tolerate the vacuum bake-out \cite{Shen2021} to achieve high vacuum. Integration of metasurfaces inside vacuum cells was experimentally realized in Ref.~\cite{Hsu2022}, where the metasurface is bonded to the vacuum cell using epoxy that survives the baking cycle (Fig.~\ref{fig:figure3}(i)). That said, operating a metasurface inside the vacuum cell next to the atoms creates more challenges. A metasurface facing the atomic source could be gradually coated by the atomic vapor, lowering the metasurface transmittance and distorting its phase profile \cite{Huckans2018}. The atoms adsorbed on the surface can also result in an electric field that can perturb the internal states of optically trapped atoms \cite{Hattermann2012}. These challenges can be mitigated using strategies from other in-vacuum optics, including heating the metasurface above the condensation point of the vapor \cite{Huckans2018}, clearing adsorbed atoms with light-induced atomic desorption \cite{Klempt2006}, separating the vapor source from the science region, and using a longer working distance to keep the optics away from the densest vapor.

\section{Materials and fabrication}

A material platform for atom-trapping metasurfaces needs to provide both sufficient phase control and low optical loss at the operating wavelength. A higher refractive index ($n$) helps achieve full $0$--$2\pi$ phase coverage at low-to-moderate aspect ratios of meta-atoms \cite{Devlin2016}, simplifying fabrication, but a high $n$ is only useful when the extinction coefficient ($\kappa$) is sufficiently small at and around the relevant visible and near-infrared (NIR) atomic transitions.

\begin{figure}[!ht]
  \centering
  \includegraphics[width=\linewidth]{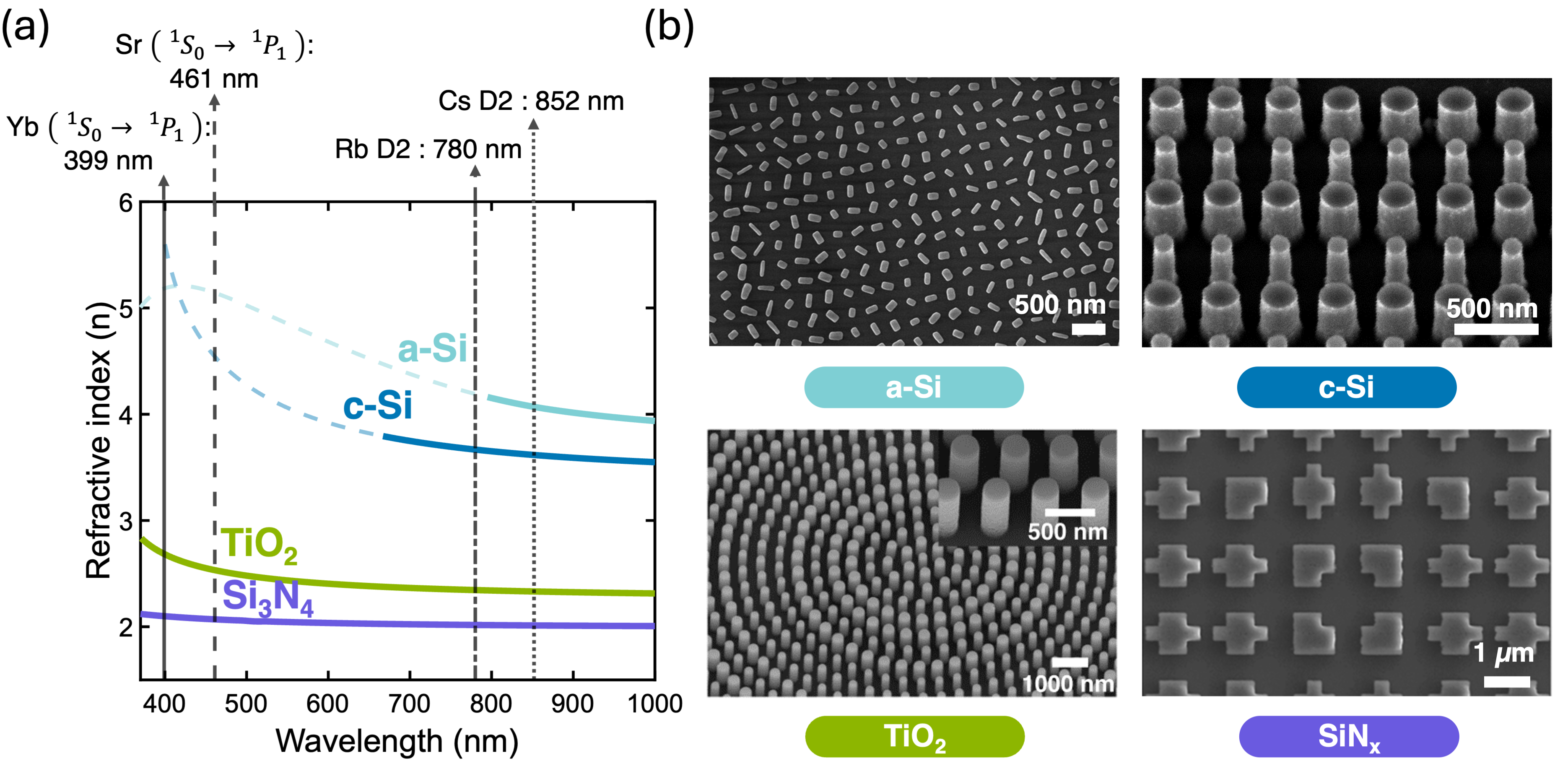}
  \caption{Optical materials for neutral-atom trapping metasurfaces (a) Refractive index ($n$) as a function of wavelength for metasurface materials, including a-Si \cite{Karaman2025}, c-Si \cite{Fang2026}, TiO$_2$ \cite{Devlin2016}, Si$_{3}$N$_{4}$ \cite{Kumar2025}. Solid curves indicate low-loss regions with $\kappa < 0.01$, while dashed curves indicate absorbing regions with a larger extinction coefficient. Vertical dashed lines mark several representative atomic transitions: Yb at 399 nm, Sr at 461 nm, Rb D2 at 780 nm, Cs D2 at 852 nm. (b) Scanning electron microscope (SEM) images of metasurfaces fabricated from a-Si \cite{Tian2025}, c-Si \cite{Fang2026}, TiO$_2$ \cite{Lim2023}, SiN$_x$ \cite{Jin2023}. SEM images adapted from \cite{Tian2025, Fang2026, Lim2023, Jin2023}.}
  \label{fig:figure4}
\end{figure}

Fig.~\ref{fig:figure4}(a) shows the refractive index of several materials that have been used for metasurfaces, along with their low-loss wavelength ranges, here defined as $\kappa < 0.01$. Due to metasurfaces being thin (i.e., the height of meta-atoms is typically hundreds of nanometers), they can tolerate significantly higher material losses than would be feasible for thick windows or for photonic integrated circuits. In some cases, the loss tolerance of metasurfaces enables the use of semiconductors for photon energy above their band gaps; for example, the extinction coefficient of crystalline silicon (c-Si) stays below 0.01 for wavelengths longer than roughly 650 nm, which is sufficiently low for metasurface designs (Fig.~\ref{fig:figure4}(a)). That being said, $\kappa < 0.01$ should be read as a rule of thumb rather than a strict cutoff for choosing materials, since the acceptable loss depends on the specific design and application. For high-power uses such as trapping large atom arrays, where the laser power can reach tens to hundreds of watts, even $\kappa = 0.01$ gives a single-digit-percent absorption that deposits a few watts of heat in a small metasurface. In such cases, an even lower-loss material is required. 

In general, there is a tradeoff between high refractive index and broader transparency window due to the inverse relationship between band gap and refractive index \cite{Moss1985}, so materials that are transparent across the entire visible and near-infrared ranges have indices no higher than $\sim$2.5. TiO$_2$ and Si have been widely used for metasurfaces \cite{Khorasaninejad2016, Liang2018, Sell2016, Zhou2017}, including for atom trapping \cite{Huang2023, Hsu2022, Chen2024, Fang2026}, though other material choices may be preferable when weighing refractive index and losses at the wavelength of interest, material and substrate availability, and fabrication complexity.

Crystalline silicon (c-Si) has a high refractive index across the NIR, where its indirect bandgap leads to sufficiently low absorption for designing metasurfaces that address Rb and Cs D-line transitions (Fig.~\ref{fig:figure4}(a)). The high index and established fabrication techniques from CMOS and silicon photonics make c-Si an attractive material for scalable metasurfaces. c-Si metasurfaces designed for atom trapping have been fabricated starting from silicon-on-insulator \cite{Huang2024} and silicon-on-sapphire \cite{Fang2026} wafers, with sapphire in particular providing a transparent, chemically inert, and UHV-compatible substrate for integration with cold-atom systems \cite{Yang2024}. Amorphous silicon (a-Si) has a narrower region of low losses because structural disorder introduces additional optical loss at most wavelengths \cite{Cody1981}, though at wavelengths $>$~800~nm it can be a good option due to its very high index and straightforward growth on transparent substrates via chemical vapor deposition (CVD) \cite{Hsu2022, Tian2025}.

Titanium dioxide (TiO$_2$) has been extensively used for low-loss visible metasurfaces, including metalenses \cite{Khorasaninejad2016, Liang2018} and reflective metasurfaces for generating colors \red{in the visible} \cite{Sun2017}. TiO$_2$ combines low absorption with a relatively high refractive index for a low-loss material in the visible ($n \approx 2.4$), enabling, e.g., high-NA metalenses \cite{Khorasaninejad2016}. For neutral-atom systems, this broad transparency range enables the use of metasurfaces to address transitions and trapping wavelengths for, e.g., Yb, Sr, Rb, Cs (Fig.~\ref{fig:figure4}(a)). High-quality TiO$_2$ metasurfaces are typically fabricated either by conformal atomic layer deposition (ALD) filling around a resist mask, followed by planarization \cite{Khorasaninejad2016, Holman2026}, or by hard-mask based etching of sputtered TiO$_2$ films \cite{Huang2023}. These approaches can produce high-quality devices, but the former is limited by the slow growth rate of ALD, whereas the latter requires challenging TiO$_2$ etching \cite{Holman2026}.

Among the material systems with the lowest losses is silicon nitride (SiN$_x$). Like Si, this material is CMOS-compatible with available foundry fabrication flows. The refractive index of SiN$_x$ is lower than that of Si but can be adjusted through composition, from stoichiometric Si$_3$N$_4$ (with the lowest index; Fig.~\ref{fig:figure4}(a)) to silicon-rich nitride \cite{Kumar2025}. SiN$_x$ metasurfaces have been used as high-NA visible lenses and vortex-beam generators with high transmission \cite{Zhan2016}. In Ref.~\cite{Holman2026}, a silicon-rich SiN$_x$ metasurface was used to generate Sr tweezer arrays at 520 nm. More broadly, SiN$_x$ is low-loss across visible-to-NIR wavelengths, which are relevant to cooling, imaging, and repumping. Due to the well-established use of SiN$_x$ in photonic integrated circuits \cite{Isichenko2023, Menon2024}, it can also be useful for hybrid platforms in which on-chip waveguides deliver and distribute light to integrated metasurfaces for beam shaping, focusing, or trap-array generation.

Even with a low-loss material, the substantial laser power needed to generate a large trap array might still deposit appreciable heat in the metasurface, making substrate choice also important for thermal management to prevent thermal damage to the metasurfaces and significant thermo-optical effects. The thermal conductivity differs significantly between common substrate materials that are optically transparent, like fused silica (near 1 W/m/K) and sapphire (tens of W/m/K). Diamond, at a thermal conductivity of $\sim$2000~W/m/K, can be an (expensive) option if much higher thermal conductivity is needed.

\section{Outlook}

We have described the advantages of metasurfaces for generating large numbers of trapping sites and complex trapping profiles, due to the straightforward scalability of metasurface pixel counts into the billions. Metasurfaces have already been used to generate arrays with 360,000 trapping sites \cite{Holman2026}, far larger than the record number of traps achieved using SLMs \cite{Manetsch2025}, and to trap about 11,000 single atoms \cite{Wang2026}, surpassing the previous record of 6,100 atoms set with SLMs \cite{Manetsch2025}. Looking forward, advanced fabrication processes such as deep-ultraviolet lithography can extend the metasurface pixel counts even more, bringing million-atom arrays into reach.

Beyond advantages in scale, the ability of metasurfaces to imprint nearly arbitrary phase, amplitude, and polarization profiles, combined with dispersion engineering of each meta-atom, enables the combination of multiple functionalities on the same optic. The integration of multiple functionalities can significantly reduce the number of optical components in atomic experiments, enabling robust and portable quantum instruments.

The downside of today's metasurfaces is that they are not readily tunable, and cannot reproduce the dynamic functionality of SLMs and AODs, for example for rearrangement of atoms or compensating for fluctuations in laser power. It is likely that near-term experiments will combine the unique advantages of metasurfaces with the dynamic tunability of SLMs and AODs, where the active components will be used to fine-tune and reconfigure atomic arrays. As an interesting alternative, atom arrays can be transported without active optical components by physically moving the optics that generate the array, such as metasurfaces and objective lenses, as proposed in ref. \cite{Xue2026}. Looking further, research into tunable pixel-addressable metasurfaces \cite{Abdelraouf2022, AbouHamdan2025} may integrate the reconfiguration functionality into the metasurfaces themselves.

\bigskip

\section*{Acknowledgments}
We thank Yuan Xu, Jiahao Wu, Nanfang Yu, and Sebastian Will for providing the photograph of the metasurface and the image of the generated trap array shown in Fig.~\ref{fig:figure1}(c).

\section*{Funding}
This material is \red{primarily} based \red{upon} work supported by the U.S. Department of Energy Office of Science National Quantum Information Science Research Centers, as part of the Q-NEXT center. Additional funding support was provided by the National Science Foundation under Award 2016136 for the QLCI center Hybrid Quantum Architectures and Networks, and by the National Science Foundation under Award 2326784 for QuSeC-TAQS.

\section*{Disclosures}
MS is affiliated with and a shareholder in Infleqtion, Inc.

\section*{Data availability}
No original data were generated for this review article.

\bibliographystyle{unsrtnat}
\bibliography{references}

\end{document}